\documentclass[aps,prd,twocolumn,groupedaddress,nofootinbib]{revtex4-1}
\pdfoutput=1
\usepackage{graphicx}
\usepackage{dcolumn}
\usepackage{bm}
\usepackage{braket}
\usepackage{verbatim}
\usepackage{amsmath}

\begin{document}

\title{Multipartite Entanglement and Firewalls}

\author{Shengqiao Luo, Henry Stoltenberg, Andreas Albrecht}

\affiliation{%
University of California at Davis; Department of Physics
\\One Shields Avenue; Davis, CA 95616
}%

\begin{abstract}
Black holes offer an exciting area to explore the nature of quantum gravity. The classic work on Hawking radiation indicates that black holes should decay via quantum effects, but our ideas about how this might work at a technical level are incomplete. Recently Almheiri-Marolf-Polchinski-Sully (AMPS) have noted an apparent paradox in reconciling fundamental properties of quantum mechanics with standard beliefs about black holes. One way to resolve the paradox is to postulate the existence of a ``firewall'' inside the black hole horizon which prevents objects from falling smoothly toward the singularity. A fundamental limitation on the behavior of quantum entanglement known as ``monogamy'' plays a key role in the AMPS argument. Our goal is to study and apply many-body entanglement theory to consider the entanglement among different parts of Hawking radiation and black holes. Using the multipartite entanglement measure called negativity, we identify an example which could change the AMPS accounting of quantum entanglement and perhaps eliminate the need for a firewall. Specifically, we constructed a toy model for black hole decay which has different entanglement behavior than that assumed by AMPS. We discuss the additional steps that would be needed to bring lessons from our toy model to our understanding of realistic black holes.

\end{abstract}

\maketitle

\section{Introduction and Background} 
Treating black holes as quantum systems have posed interesting questions about what the fundamental properties of quantum mechanics should be in the context of gravity. The so-called black hole information paradox is the apparent contradiction during the evaporation of black holes through emission of Hawking radiation between the causal structure inherent to black holes and the overall unitarity of quantum mechanics\cite{Hawking:1976ra}. As an illustration, consider a black hole that begun as an initially pure state\footnote{This is a common assumption when stating the problem but is done for simplicity. Relaxing this assumption does not resolve the information problem}. Unitary quantum evolution requires that the evolving state should remain pure at later times assuming no interactions with external systems are present. Inspired by the ideas of Bekenstein and Hawking, black holes can be thought of as thermodynamic systems with a temperature and entropy\cite{Bekenstein:1973ur, Hawking:1974sw}. They will radiate like a black body and can decrease in size and mass and possibly eventually evaporate away completely. If we consider an initially pure black hole evolving completely into Hawking radiation, then the unitarity of quantum mechanics would seem to imply that the final Hawking radiation state should be pure. However, the process of creating Hawking radiation should be analogous to pair production due to quantum fluctuations. This would seem to suggest the radiation produced should be mixed and due to the casual structure of the black hole, there wouldn't be interactions between radiation emitted early and very late in the black hole's history capable of undoing the mixed property.

Black hole complementarity proposes a resolution of the black hole information paradox which allows a pure final state. The idea is that different causally disconnected observers view different yet complementary pictures which disagree on the location of the information encoded in the matter that created the black hole \cite{Susskind:1993if}. Since the observers could not communicate, a contradiction would not be seen by any one observer. In recent years, the debate has shifted; Almheriri, Marolf, Polchinski and Sully (AMPS) presented a simple argument that revealed\footnote{Prior to this, Braunstein et al. came to similar conclusions and described the deviation from vacuum near the horizon as an ``energetic curtain" \cite{Braunstein:2009my}.} approaches such as complementarity do not seem to be enough and that a contradiction between the two complementary pictures could be observed \cite{Almheiri2013}. They suggested that the most modest resolution would be preventing information from entering the interior of the black hole during late times by introducing a high energy barrier called a firewall.

The quantum information present in a black hole and its decay products can be explored by studying their entanglement. As demonstrated by Page \cite{Page1993}, to achieve a final pure state of Hawking radiation requires entanglement between early and late radiation in this final state. This late time entanglement constrains the type and level of entanglement present throughout the evaporation process. In particular, AMPS argues that entanglement for late radiation with Hawking modes behind the horizon seems to be forbidden which would result in a breakdown of the field theory vacuum \cite{Almheiri2013}. This argument evokes a well known basic property of quantum mechanics called quantum monogamy which constrains how entanglement can be shared.

Our paper focuses on the role quantum monogamy plays in this problem. In the case of maximal bipartite type entanglement (such as the entanglement present in a Bell pair), quantum monogamy gives a simple conclusion. If system A is maximally entangled with system B then no entanglement can exist between system A and any third system. When deviating from maximal bipartite entanglement, monogamy inequalities limit entanglement with a third system. We use a toy model which deviates from maximal bipartite entanglement for an evaporating black hole to illustrate that we can impose an entanglement structure for early and late times and show entanglement across the horizon is not strictly forbidden. Since any realistic physical system will not absolutely saturate maximal entanglement, our exploration is motivated by the possibility that even extremely small deviations from maximal entanglement for realistic black holes could lead to conclusions very different from those of AMPS.  

In this paper we consider the information that originated in the initial state of the black hole becoming encoded in multipartite entanglement. Other authors \cite{Braunstein:2009gv, Braunstein:2006sj} have stated that tripartite type entanglement naturally arises from black hole evaporation. There are important differences between our work and theirs. The conclusions in \cite{Braunstein:2009gv, Braunstein:2006sj} are a consequence of how their Hilbert Spaces are constructed and partitioned which differs from what we do in this paper.  In particular the difference is due to the nature of their ”external neighborhood” with which the black hole interior remains entangled. Their construction presents a resolution to the information paradox but leads to a firewall-like conclusion. Our different division of Hilbert spaces and construction of black hole and radiation states allow us to make different assumptions about the quantum mechanics of black holes and leads to different conclusions about the nature of the near horizon region. Our work also differs from \cite{Stoltenberg2015, Hwang:2016otg} which involve a separate third system to entangle with while we consider entanglements between radiation subsystems and the black hole without the need of an external system. We argue that our approach suggests a way to avoid firewalls.

Our paper is organized as follows: Section \ref{sec:Postulates} presents our list of beliefs that we assume to be true of the nature of an evaporating black hole.  Section \ref{sec:Multipart} introduces some basic facts about multipartite entanglement, and contrasts multipartite vs. bipartite entanglement.  Section \ref{sec:toymodelsetup} sets up our toy model, referencing a toy model from earlier work \cite{Stoltenberg2015} \footnote{A different approach to a qubit toy model can be found in \cite{Osuga:2016htn}. Their work avoids the firewall argument with different assumptions of the form of black hole states and the division of Hilbert spaces.} designed to illustrate the AMPS argument. Section \ref{sec:Results} gives the results of calculating both bipartite and multipartite entanglement measures for the toy model and mentions challenges with interpreting these numbers.   Section \ref{sec:Discussion} discusses the interpretation of our toy model in the context of the firewall problem.  Section \ref{sec:Vacuum} examines how our results fit into current understanding of the field theory vacuum. We present conclusions in section \ref{sec:Conclusions}, and various technical results in the appendices.

\section{Principles of Black Hole Evolution} \label{sec:Postulates}
How the black hole information problem is resolved or remains a problem boils down to what you assume is true of black hole evolution. In this section, we describe what we assume for evaporating black holes and how these beliefs are realized in our toy model.
Our assumptions of what black hole evolution looks like far from the horizon mirrors postulates of black hole complementarity presented in \cite{Susskind:1993if}.

Postulate 1: The process of formation and evaporation of a black hole, as viewed by a distant observer, can be described entirely within the context of a standard quantum theory. In particular, there exists a unitary S-matrix which describes evolution from in-falling matter to outgoing Hawking-like radiation.

Evolution in our model is explicitly unitary which is achieved by preserving the purity of the initial state of the entire system as the black hole evaporates.

Postulate 2: Outside the stretched horizon of a massive black hole, physics can be described to good approximation by a set of semi-classical field equations.

Our toy-model constructed out of qubits is too simple to check if it is consistent with this assumption. There does not seem to be anything that directly conflicts with it though, so we assume that our model does not present any contradictions here.

Postulate 3: To a distant observer, a black hole appears to be a quantum system with discrete energy levels. The dimension of the subspace of states describing a black hole of mass $M$ is the exponential of the Bekenstein entropy $S(M)$.

We model this with finite Hilbert spaces to describe our evolving black hole system. As it evaporates, the dimension of the subspace of states decreases to coincide with decreasing mass.

Unlike the exterior, The nature of a black hole's interior has not yet been observed. We expect and require the nature of the black hole's interior to be consistent with the assumptions of the exterior that we have listed above. For our approach, we suggest some very non-local behavior occurring in the interior. We expect the standard effective field theory description of pair production near the horizon to fail somewhere (but not necessarily in a way that creates a firewall). We require the horizon to retain some memory of the black hole's history. In order to enforce the proper type of late time entanglement, we wish to transfer the entanglement from the infalling partners to entanglement with the horizon. In our toy model, our black hole interior states do not have a description with any causal or spatial structure. We expect that an extension of our ideas to a more realistic theory with a spatial interpretation of the interior would include very non-local behavior in the interior.

\section{Multipartite entanglement} \label{sec:Multipart}
In this section we present a general discussion of multipartite entanglement.  We will use the concepts presented here to motivate and analyze the toy model presented in ~Section \ref{sec:toymodelsetup}. 

A common notion of entanglement is defined by inseparability of states. First consider separability for pure states. A pure state living in a product space with a bipartition, $A \otimes B$ can always be written using the Schmidt decomposition as:
\begin{eqnarray} \label{eq:pure1}
\ket{\Psi_{A \otimes B}}= \sum_{i}c_i\ket{i_{A}}\otimes\ket{i_{B}}
\end{eqnarray}
with bases $\ket{i_A}$ and $\ket{i_B}$ for Hilbert spaces $A$ and $B$ respectively and coefficients $c_i$. A state is separable only if the state can be written as a single term in this sum:
\begin{eqnarray} \label{eq:pure2}
\ket{\Psi_{A \otimes B}}= \ket{\psi_{A}}\otimes\ket{\phi_{B}}
\end{eqnarray}
for some $\psi_{A}$ and $\phi_{B}$ living in Hilbert spaces $A$ and $B$ respectively. The state, $\ket{\Psi_{A \otimes B}}$ being separable is completely equivalent to the Von Neumann entropy after tracing out system $A$ or $B$ being zero. Otherwise it is inseparable or in other words entangled. For this reason Von Neumann entropy serves as measure of entanglement in pure states with bipartitions. 

More generally, a state described by a density matrix, $\rho_{A \otimes B}$ which describes either a mixed or pure state is separable only if it can be written as:
\begin{equation} \label{eq:mixed1}
\rho_{A \otimes B} = \sum_{i} p_i \rho_{A,i} \otimes \rho_{B,i}
\end{equation}
for some collection of density matrices, $\rho_{A,i}$ for subsystem $A$ and $\rho_{B,i}$ for $B$ and coefficients $p_i$ with $\sum_i{p_i}=1$.\footnote{Note that the definition of separability for pure states is consistent with the more general definition of separability.} If there exists no decomposition of this form for a state then that state is inseparable and is therefore entangled. Since there does not exist a sufficiently analogous decomposition to Schmidt decomposition for mixed states, separability can be much harder to check. Also, Von Neumann entropy is known not to be a useful indicator of entanglement between two systems which together are in a mixed state.

To understand the entanglement between systems that are parts of an overall mixed state, negativity \cite{Vidal2002}, entanglement of formation\cite{Wootters1998}, distillable entanglement\cite{1996PhRvL..76..722B, 1999PhRvA..60..173R}, and concurrence\cite{Wootters1998} are useful quantities to be considered. Among all these entanglement measures, negativity has the benefit of being generally calculable. In this paper we will focus on calculating negativities in our toy models. The technical definition and properties of negativity can be found in Appendix \ref{sec:negativity}. The main property of negativity we are interested in is that if the negativity between two subsystems is non-zero then the subsystems are inseparable\footnote{The converse of this statement is not true. Entangled subsystems can have zero negativity.} i.e. entangled. 

To illustrate some interesting multipartite entanglement properties states can have, consider the $W$\cite{Dur2000, Nielsen1999} and $GHZ$\cite{Greenberger2007, Dur2000} states. The $W$ state, when made up of three qubits has the form:
\begin{equation}\label{eq:wstate}
W = \frac{1}{\sqrt{3}}(\ket{\downarrow\uparrow\uparrow}+\ket{\uparrow\downarrow\uparrow}+\ket{\uparrow\uparrow\downarrow}).
\end{equation}
For this state, tracing out any system leaves a mixed state. The nonzero Von Neumann entropy for any single qubit directly implies entanglement between any qubit and the remaining two quits but fails to reveal entanglement between any two qubits (because any two qubits together are always in mixed state). The interesting property of this state is its maximal entanglement ``robustness"\cite{1999PhRvA..59..141V} meaning it retains the most entanglement after ``disposal" (tracing out) of one qubit for three qubits systems.
After tracing out one of the qubits, the negativity\cite{Vidal2002} between the remaining two qubits is non-zero. This means one qubit in the W state concurrently shares entanglement with each other qubit individually.

Compare this to the $GHZ$ state:
\begin{equation}\label{eq:ghzstate}
GHZ = \frac{1}{\sqrt{2}}(\ket{\uparrow\uparrow\uparrow}+\ket{\downarrow\downarrow\downarrow}).
\end{equation}
Like the $W$ state, for the $GHZ$ state each qubit exists in a mixed state. However unlike the $W$ state, after tracing out any one qubit, the negativity of the remaining two qubits is zero. Furthermore, after tracing out a qubit, you are left with a density matrix in the form of ~Eqn. \ref{eq:mixed1} meaning the remaining two qubits are also completely separable. In other words, when any one qubit is traced out, no entanglement remains between the other two qubits; In the $GHZ$ state, entanglement exists between each single qubit and the remaining pair of qubits but no entanglement exists between any pair of qubits.

States like these motivated us to study shared entanglement structures in the context of the black hole information problem. We use negativity to reveal the existence of entanglement between particular subsystems for a toy model of an evaporating black hole and its radiation.

\section{Toy Model Setup}\label{sec:toymodelsetup}

\begin{figure} \label{fig:bits}
	\includegraphics[width=9cm,height=4cm]{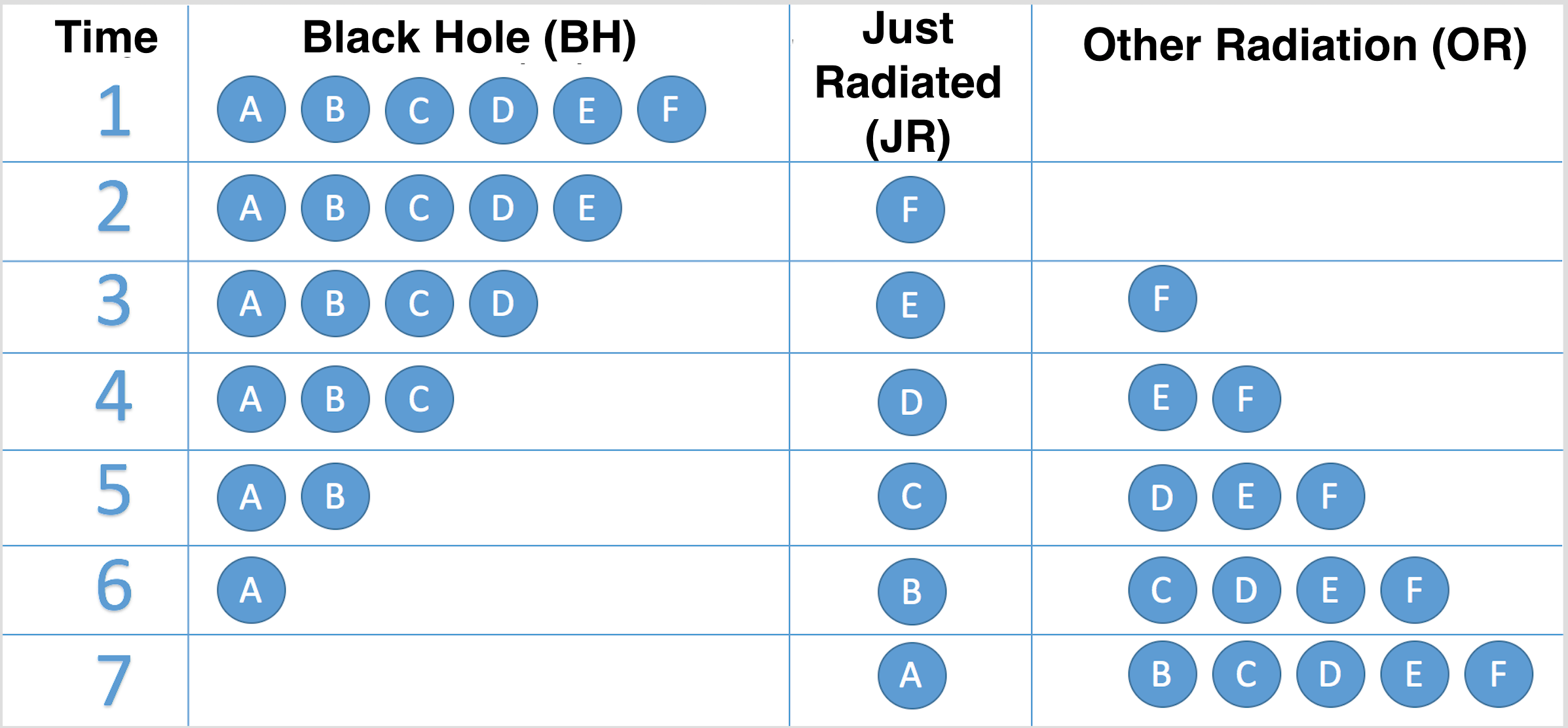}
	\caption{For each time step we illustrate the allocation of our six qbits (labeled A-F) among the three physical roles:  i) Part of the Black Hole, ii) Just radiated Hawking radiation, and iii) Other Hawking radiation.  As time goes on the allocation shifts, corresponding to the decay process of the black hole.}
\end{figure}

We will use a simple toy model similar to the ones used in \cite{Stoltenberg2015}. In this section we will first describe how the Hilbert spaces are divided and quantum states are described, then we will describe the time evolution and follow up with specific information about the models. The first toy model in \cite{Stoltenberg2015} illustrates the black hole information problem and firewall argument by describing the entanglement with collections of Bell pairs of qubits. Here we will refer to that toy model as the Firewall toy model. We compare this to an analogous toy model created for this paper with a different entanglement structure which we will refer to as the Multipartite toy model.  For both models, the black hole and all the radiation are represented by six qubits. We keep the size of the entire Hilbert space constant, always dimension $2^6$. At each time step we reassign the degrees of freedom of the Hilbert space to either black hole degrees of freedom or Hawking radiation degrees of freedom. We begin the black hole in a pure state and describe the evaporation by giving the toy model state at seven specific points in time.\footnote{We imagine our system is undergoing some continuous evolution but we are examining the system only at 7 discrete (not evenly spaced) time steps.} Each time step describes one Hawking qubit being emitted from the black hole.  For our analysis at each time step we will partition the system into 3 parts: black hole (BH), just emitted particle (JR) and other radiation (OR) and examine the entanglement between them. An illustration of the assigning of qubits to either describe the black hole or the decay products at each time step is shown in ~Fig. \ref{fig:bits}.

The Firewall toy model explicitly demonstrates the conflict of trying to simultaneously enforce overall unitary time evolution (which results in entanglement between early and late radiation) and trying to enforce ``no drama" at the horizon (which would require entanglement across the horizon region as seen by an infalling observer). These two properties seem to be in conflict because of quantum monogamy and no cloning theorems which are key properties of quantum mechanics. If a qubit is maximally entangled with another qubit (thus forming a Bell pair), then neither qubit can share any entanglement with another system. In the Firewall toy model examining the bipartite entanglement suffices to demonstrate the conflict.

In the Multipartite toy model, we do not force the qubits to appear in parts of Bell pairs. We allow the newly emitted Hawking qubit to share multipartite entanglement with the black hole system and radiation system. Multipartite entanglement offers a richer structure of entanglement sharing and we will see the conclusions of forbidden entanglement across the horizon for late times no longer holds.

The full $2^6$ dimensional Hilbert space, $U$ is described by six qubits (labeled $A$ through $F$) as:
\begin{equation}\label{eq:hspace}
U = A \otimes B \otimes C \otimes D \otimes E \otimes F. 
\end{equation}
As the black hole evaporates, qubits are reassigned from describing black hole degrees of freedom to radiation degrees of freedom (they are emitted in reverse alphabetical order). At each time step, our tripartition looks like:
\begin{equation}\label{eq:hspace2}
U = H_{BH} \otimes H_{JR} \otimes H_{OR}
\end{equation}
using our subdivision into Black Hole ($BH$), Just Radiated ($JR$) and Other Radiation ($OR$). The specific qubits that describe each subsystem changes with each time step (as shown in Fig. \ref{fig:bits}). 

To get an idea of what the states look like, consider an example state in the Firewall toy model at time 3 (two particles have radiated away)\footnote{The time steps and labels used in this paper differ than those used in \cite{Stoltenberg2015}. In this paper we have chosen the Firewall and Multipartite models to have consistent labels with each other.}. The state of the system at this time contains Bell pair-like bipartite entanglement between the $JR$ particle and $BH$:

\begin{eqnarray} \label{eq:time3}
\frac{1}{2}&[\left(\begin{array}{ccc}
&\ket{0_{JR}0_{BH}} \\
+&\ket{1_{JR}1_{BH}} \end{array} \right) \ket{0_{OR}} + \\
&\left( \begin{array}{ccc}
&\ket{0_{JR}2_{BH}} \\
+&\ket{1_{JR}3_{BH}} \end{array} \right) \ket{1_{OR}}]
\end{eqnarray}
In these states, the numbers in the kets enumerate the basis vectors of the subsystem. Here we can explicitly see entanglement between the $JR$ and $BH$ subsystems. Entanglement between the $OR$ and $JR \otimes BH$ is also present but due to the symmetry in the $JR$ elements, no entanglement between $JR$ and $OR$ exists (they are completely separable).

In the Multipartite toy model, we impose a more complicated entanglement structure at time 3:
\begin{eqnarray}\label{eq:newstruc}
&a(\ket{0_{OR}1_{JR}}+\ket{1_{OR}0_{JR}})(\ket{0_{BH}}+\ket{1_{BH}}) \\ \nonumber
&+b(\ket{0_{JR}1_{BH}}+\ket{1_{JR}0_{BH}})(\ket{0_{OR}}+\ket{1_{OR}})
\end{eqnarray}
Here we write the state as a linear combination of states where the $JR$ system is maximally entangled with the $OR$ system and a state where the $JR$ system maximally entangled with the $BH$ system. The combination comes with arbitrary (up to normalization) coefficients $a$ and $b$. Here as long as neither coefficient is $0$, there exists some level of shared entanglement between the three systems (under certain measures). In this case there does not exist maximal bipartite type entanglement between any two of these three subsystems.

In both toy models, the $JR$ qubit comes out maximally mixed (with Von Neumann entropy 1). In the firewall toy model, a $JR$ qubit comes out completely entangled with the $BH$ for early times and for late times switches to coming out completely entangled with $OR$. In the multipartite toy model, at every time after time step 3, the emitted JR qubits come out with shared entanglement between $OR$ and $BH$.

At the core of the black hole information problem, is difficulty trying to enforce overall unitary evolution for black hole evaporation. Fundamentally, unitary evolution is just an inner product preserving and onto mapping of states. Within each toy model, we only examine one example evolution for some chosen initial state. Our overall evolution looks something like: 
\begin{equation}\label{eq:timeevo}
U(t): \Psi_{t1} \rightarrow \Psi_{t2}\rightarrow \Psi_{t3}\rightarrow...
\end{equation}
In principle, we could have created a Hamiltonian to generate such time evolution. In this work, we are not interested in the form of the Hamiltonian but instead the entanglement properties of the states as the system evolves. Violating unitary evolution would involve evolving an overall pure state into a mixed state. We enforce unitarity by writing a list of states that remain pure for the entire history of our system.

From the perspective of a stationary observer outside the horizon, we require the evaporation process to only involve local unitary interactions. We will take that to mean that quantum entanglement can only change when a Hawking quantum is just leaving the horizon due to interactions that occur there. This means that new entanglement can be generated between $BH$ and $JR$ as a new Hawking quantum is created. Also, the entanglement between the $BH$ and $JR$ systems can transfer to entanglement between $BH$ and $OR$ and entanglement between $OR$ and $BH$ can transfer to entanglement between $OR$ and $JR$ due to the shifting of quanta from one category to another (specifically, quanta used to describe $BH$ becoming $JR$ quanta and $JR$ quanta becoming $OR$ quanta). In our toy model, the density matrices of every combination of previously emitted radiation (i.e. all radiation other than the just emitted particle) are forced to remain unchanged for all subsequent time steps. This restriction is somewhat stronger than is absolutely necessary since the only important constraint is that the eigenvalues of these density matrices do not change. We treat the black hole in this toy model as a ``black box''. As the system evolves, we do not explicitly know what the quantum nature of the black hole truly is other than some very coarse grained properties such as the Hilbert space dimension it lives in and its Von Neumann entropy. There in principle could be some very non-local behavior in the interior. 

To see how this works explicitly, we will first illustrate an example of going from time step 3 to time step 4 in the Firewall toy model (which is easier to eyeball than the Multipartite toy model). Starting with our system at time 3, in the state described by Eqn. \ref{eq:time3}, we evolve to time state 4, by shifting the qubits used to describe each system. Going from time 3 to time 4, the size of the black hole has decreased and the number of radiated particles has increased. We model that by having the states that described the black hole ($BH,3$) at time step 3 become states that describe the black hole ($BH,4$) and the newly emitted particle ($JR,4$) at time step 4. This is done by writing the basis for ($BH,3$) in terms of ($BH,4$) and ($JR,4$):
\begin{eqnarray} \label{eq:map1}
\ket{0_{BH,3}}=\frac{1}{\sqrt{2}}(\ket{0_{{JR,4}}0_{{BH,4}}}+\ket{1_{{JR,4}}1_{{BH,4}}}) \nonumber \\ 
\ket{1_{BH,3}}=\frac{1}{\sqrt{2}}(\ket{0_{{JR,4}}2_{{BH,4}}}+\ket{1_{{JR,4}}3_{{BH,4}}})  \\
\ket{2_{BH,3}}=\frac{1}{\sqrt{2}}(\ket{0_{{JR,4}}4_{{BH,4}}}+\ket{1_{{JR,4}}5_{{BH,4}}}) \nonumber \\
\ket{3_{BH,3}}=\frac{1}{\sqrt{2}}(\ket{0_{{JR,4}}6_{{BH,4}}}+\ket{1_{{JR,4}}7_{{BH,4}}}) \nonumber 
\end{eqnarray}
Here we include an additional subscript label, which enumerates the time step (in this case for time $3$ and time $4$). The change of basis written above prescribes what the next time state will be, based on the previous time state. The choice of basis change we have made is written in a convenient basis and results in a particular entanglement structure. Making different choices of this basis reassignment is effectively implementing different types of interactions at the horizon and results in differing resulting entanglements.

The remaining parts of the system at time 3, ($JR,3$) and ($OR,3$) will also need to have a qubit reassigned to ($OR,4$) at time 4 as Eqn.~\ref{eq:map2}:
\begin{eqnarray} \label{eq:map2}
\ket{0_{JR,3}0_{OR,3}} = \ket{0_{OR,4}} \nonumber \\ 
\ket{0_{JR,3}1_{OR,3}} = \ket{1_{OR,4}} \\ 
\ket{1_{JR,3}0_{OR,3}} = \ket{2_{OR,4}} \nonumber \\ 
\ket{1_{JR,3}1_{OR,3}} = \ket{3_{OR,4}} \nonumber 
\end{eqnarray}
This mapping should be thought of differently than Eqn.~\ref{eq:map1}. The particle emitted at time step 3 is lumped into the ($OR,4$) subsystem by time step 4. There are no interactions here (and no change in entropy) so this evolution can be thought of as just relabeling. In general we could have evolved the state through trivial phase rotations without introducing interactions but for simplicity we freeze the evolution when no more interactions occur.

We plug in these changes of basis given by ~Eqn.\ref{eq:map1} and ~Eqn.\ref{eq:map2} into the state we had previously written for time 3, (~Eqn.\ref{eq:time3}) and obtain the entire state for time 4:
\begin{eqnarray} \label{eq:map3}
\frac{1}{2\sqrt{2}}[\left( \begin{array}{ccc}
&\ket{0_{JR}0_{BH}} \\
+&\ket{1_{JR}1_{BH}} \end{array} \right) \ket{0_{OR}} \\+ 
\left( \begin{array}{ccc}
&\ket{0_{JR}2_{BH}} \\
+&\ket{1_{JR}3_{BH}} \end{array} \right) \ket{1_{OR}}  \\
+\left( \begin{array}{ccc} 
&\ket{0_{JR}4_{BH}} \\
+&\ket{1_{JR}5_{BH}} \end{array} \right) \ket{2_{OR}} \\+ 
\left( \begin{array}{ccc}
&\ket{0_{JR}6_{BH}} \\
+&\ket{1_{JR}7_{BH}} \end{array} \right) \ket{3_{OR}}] \nonumber 
\end{eqnarray}
The overall process is unitary since the entire state remains pure. 

To reiterate, we realize locality by only allowing interactions to occur as the just emitted particle leaves the horizon. During early time evolution, entanglement is created between the black hole and the just radiated qubit. During the late time evolution, the newly emitted qubits become entangled with the earlier radiation. This does not occur due to interactions with earlier radiation but instead relabeling black hole degrees of freedom as new radiation degrees of freedom. In other words, what was previously entanglement between earlier radiation and the black hole becomes entanglement between earlier radiation and new radiation.

Now that we have described the rules for how future time states are constructed, we will now describe key features of the Firewall and Multipartite models. Both models begin with a pure quantum state only describing a black hole. Each time a state describing the next time is generated in the Firewall model, the basis change for the $BH$ system takes the form of Eqn. \ref{eq:map1}. Each basis vector of $BH$ is mapped onto a state that takes the form of a Bell pair between the new $JR$ particle and a subspace of the new smaller $BH$ system. This evolution ensures that the final state at the end of evaporation is entirely comprised of pairs of quanta in Bell-pairs, each exhibiting maximal bipartite type entanglement. Our Multipartite model does not have this constraint imposed on its evolution. The most general form that for remapping of basis vectors, $\ket{k_{BH,3}}$ can take are: 
\begin{eqnarray}
\ket{k_{BH,3}}=\sum_{i,j}a_{i,j}^k\ket{i_{{JR,4}}j_{{BH,4}}}
\end{eqnarray}
with the $a_{i,j}^k$ coefficients chosen such that the basis vectors remain orthonormal. Particular choices for these coefficients will result in non-trivial multipartite entanglement structure in the final state of radiation. 

A detailed list of our states for the time evolution can be found in Appendix \ref{sec:time evo}. Even for this small and simple system, by late times the form of the states grow increasingly complex and it becomes difficult to intuitively see the entanglement. When choosing these states we used a combination of intuition, trial and error and exploring the space of coefficients with a computer. We recommend that the reader first consider the general analysis of the properties of our toy model states presented in the following section before looking at the details presented in the Appendix \ref{sec:time evo}.

\section{Results} \label{sec:Results}

The measures we use in this paper are Von Neumann entropy and negativity, each of which measures different aspects of entanglement. First we calculated Von Neumann entropies for the $BH$, $JR$ and $OR$ systems through the black hole's evaporation in Table~\ref{table:1}. The dashes in Table \ref{table:1} (as well as Tables \ref{table:2} and \ref{table:3}) imply non-existing Von Neumann entropy values since the subsystems aren't defined at those times. For example, in the first time, since there is no subsystem JR and OR, thus Von Neumann entropy of JR and OR are not well defined.

The upper bound on Von Neumann entropy
is given by:
\begin{equation} \label{eq:upbound}
0 \leq S(\rho) \leq log_{2}(D)
\end{equation}
where $D$ is either the dimension of the measured subsystem or the rest of the space which forms a purification with the measured system, whichever is smaller.

\begin{table}[h!]
	\centering
	\begin{tabular}{ |p{0.8cm}||p{2.5cm}|p{2.5cm}|p{2.5cm}|  }
		\hline
		\multicolumn{4}{|c|}{Von Neumann entropies in the Multipartite Model} \\
		\hline
		Time & $S(\rho_{BH})$ &$S(\rho_{JR})$&$S(\rho_{OR})$\\
		\hline
		1 & 0.00 & - & -\\
		2 & 1.00 & 1.00 & -\\
		3 & 2.00 & 1.00 & 1.00\\
		4 & 2.01 & 1.00 & 2.00\\
		5 & 1.32 & 1.00 & 2.01\\
		6 & 0.60 & 1.00 & 1.32\\
		7 & - & - & -\\
		\hline
	\end{tabular}
	\caption{Von Neumann entropy of three subsystems compared to the maximum possible value of Von Neumann entropy in our Multipartite model at each time step. The dashes imply non-existing Von Neumann entropy values since the subsystems aren't defined at those times.  The values from the first column of this table are plotted as $x$-s in ~Fig. \ref{fig:entropy}.}
	\label{table:1}
\end{table}
\begin{figure} \label{fig:entropy}
	\includegraphics[width=9cm]{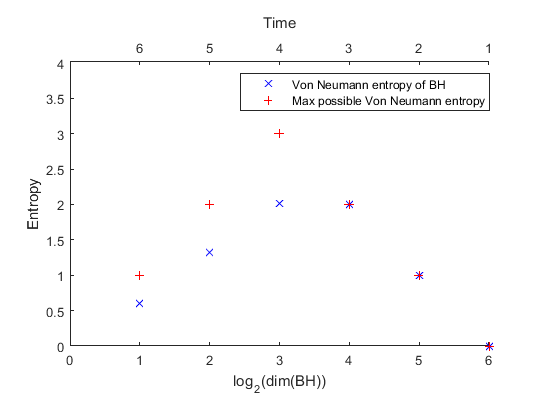}
	\caption{Calculated Von Neumann entropy of black hole sub-system (`x'-s) compared to the maximum possible value of Von Neumann entropy in our multipartite model (`+'-s) at each time step. The maximum values of entropy are based on the dimension of the smaller Hilbert Space between the measured subsystem and the rest of the space which together form a puriﬁcation.  This plot exhibits standard behaviors discussed by Page\cite{Page1993a}.}
\end{figure}
For every time, when a new Hawking qubit is radiated (subsystem $JR$), the emerging qubit is always maximally mixed, with the rest of our system serving as its purification. This can be seen from ~Table \ref{table:1}, where the Von Neumann entropy for subsystem $JR$ always numerically saturates the maximum ($\log_2 (2) = 1$) for a Hilbert space of dimension 2. We enforced this property when choosing states for our toy model since this reflects what we expect from the pair production process.

Fig. \ref{fig:entropy} shows that the entropy of $BH$ grows until the Page Time (time 4 in our system), after which it decays until the end of evaporation. This mirrors the well known curve found by Page in \cite{Page1993}. Von Neumann entropy demonstrates how mixed each subsystem is at each time but since combinations of subsystem are mixed, the entropy will not state which subsystem is entangled with which. The question of whether or not a firewall exists depends on entanglement. Here we consider multipartite entanglement across the horizon region which Von Neumann entropy of $BH$ does not reveal but negativity can.

In table \ref{table:2} we calculate negativity for our Multipartite model.

\begin{table}[htp!]
\centering
\begin{tabular}{ |p{2cm}||p{2cm}|p{2cm}|p{2cm}|  }
 \hline
 \multicolumn{4}{|c|}{Negativities in the Multipartite Model} \\
 \hline
 Time & BH and JR &JR and OR&BH and OR\\
 \hline
		1 & - & - & -\\
		2 & 0.500 & - & -\\
		3 & 0.500 & 0.000 & 0.500\\
		4 & 0.023 & 0.000 & 0.545\\
		5 & 0.038 & 0.000 & 0.583\\
		6 & 0.001 & 0.424 & 0.233\\
		7 & - & - & -\\
 \hline
\end{tabular}
\caption{Negativity values between all combinations of our three subsystems $BH$, $JR$ and $OR$ for all times in our Multipartite toy model. Negativity between BH and JR are nonzero for times in which negativity values are well defined. This implies entanglement across the horizon is not strongly forbidden. }
\label{table:2}
\end{table}
At all times including the Page time, when negativity values between $BH$ and $JR$ are defined, negativity between $BH$ and $JR$ is nonzero. 
As it stated earlier, nonzero values of negativity always means the subsystems are inseparable. Our main result is that we find entanglement between $BH$ and $JR$ for the entire history of evaporation. This differs from the AMPS's expectation that after the Page time, entanglement between $BH$ and $JR$ should be forbidden. Further intuition about negativity is developed in ~Section \ref{sec: firewall model table}, relating our results to \cite{Braunstein:2009my} and comparing negativity values between the Multipartite and Firewall toy models.

However there are limits to what negativity can teach us. Because no upper bound for negativity in a mixed state is known, the specific meaning of a finite nonzero value is unclear. Therefore, attempting to compare negativity values from different time steps can be misleading. The changing dimensionality of subsystems for different time steps surely adds to this uncertainty. Additionally, special states called ``PPT"states have entanglement even though they have negativity equal to zero\cite{Vidal2002}.

A good way to move forward despite these confusing aspects of negativity would be to compare the multipartite entanglement existing across the horizon in our toy model with the entanglement in a realistic field theory that could give some physically meaningful point of reference. In the remainder of this paper we explore what such a comparison would entail.

\section{Discussion} \label{sec:Discussion}
Answering the question of whether or not there is a firewall involves asking what is seen by an observer falling into the black hole. The classical GR result is that an infalling observer would not see the horizon as a special location, looking no different than flat space which is in part a consequence of the equivalence principle. Naively, a proper treatment of quantum gravity would not seem to change this result since for suitably large black holes the energy scale set by the curvature near the horizon is far below the Planck scale and the usual expectation is that classical GR should apply. However, AMPS would argue that any attempt at black hole complementarity would fail at late times and no vacuum-like field theory description can be found in the interior of the black hole due to entanglement that already exists outside the horizon between early and late radiation preventing entanglement across the horizon. In our Multipartite toy model, we have created an example for the time evolution where an infalling observer can find entanglement across the horizon and a complementary description could possibly exist. This is however a single example for the time evolution, and without more knowledge of how black hole evaporation should look on the quantum level, we cannot make any claims that this evolution would be typical \footnote{The firewall argument can change with different assumptions about typical states. An example of this can be found in \cite{Hossenfelder:2014jha} which unlike this paper, considers an evolution that gives states with an entanglement structure that deviates from the Page result. Also, in the $ER=EPR$ proposal, it has been argued that typical states do not have firewalls\cite{Susskind:2015toa}}.

Furthermore, we don't know how the results from our toy model generalize to an actual field theory description. A well-known property of the vacuum in field theory is its purity and the presence of entanglement. Although our work does not address restoring the purity\footnote{A possible extension to our work would be to create a complementarity scheme that explicitly restores purity such as taking states from or analogous to those in our multipartite model and remapping them for an infalling observer to construct a pure vacuum state.} that is typically assumed for the vacuum state, we have considered a new entanglement structure that differs from the firewall literature. Our multipartite model does not offer maximal bipartite type entanglement between quanta leaving the horizon and those falling in. To see how this could differ from the field theory vacuum state, we would want to know how our proposal would change the energy per field mode vs. the ground state. Even a small change per mode could give a divergent total effect, which would restore the original arguments of AMPS \footnote{We thank Steven Carlip for bringing up this point to us.}. Answering these questions is crucial for our ideas to be important to the actual black hole firewall problem and requires more technical work than we offer at this point. Still, we offer some further reflections on this issue.

\section{Field Theory Vacuum Entanglement} \label{sec:Vacuum}

In this section, we discuss how our results fit into current understanding of vacuum entanglement. First, for an operational description of the nature of entanglement in ground states, we will state some well known results in flat space. If an observer falling into a black hole did actually encounter the vacuum (or some close approximation) in the horizon region, then we would expect the same properties to be present. First is the presence of entanglement in Unruh radiation. Unruh radiation is analogous to Hawking radiation for accelerating observers in flat space. If two uniformly accelerated observers accelerated in opposite directions with the same acceleration, they would encounter pairs of entangled particles. These observers could generate entanglement between each other despite being outside of each other's lightcones and therefore not causally connected. The interpretation is that the observers are harnessing already existing entanglement from the vacuum state in flat space. Through this construction, a pair of observers could create Bell pairs which could suggest Bell pair-like entanglement could be inherent to the vacuum. Also, \cite{PhysRevA.71.054301,PhysRevA.90.032316} have proposed various experiments which could extract multipartite type entanglement structures (like GHZ states) analogous to the Bell pair extraction process just described. These appear to make the case for a more complicated multipartite entanglement structure of the vacuum.  However, because we have not made a clear connection between our toy model and field theory, we are unable to argue that the multipartite entanglement considered in \cite{PhysRevA.71.054301,PhysRevA.90.032316} is ``the same” as that exhibited in our toy model. 

Entanglement in the ground state is also revealed using the path integral formulation for quantum field theory and is applicable to CFT's. In this formulation density matrices can be constructed and represented as path integrals. When tracing out regions of space, the remaining density matrix will have a Von Neumann entropy that scales with the area dividing the region \cite{Nishioka2009}. As we have stated earlier, Von Neumann entropy is a useful measure of mixedness of a state as well as measuring entanglement across a bipartition for an overall pure state, but does not reveal all aspects of entanglement. Calculating other measures of entanglement in this formulation is much more difficult although negativities have been calculated in 1+1D CFT's\cite{Rangamani2014}.

In all of these constructions, it is difficult to fully explore all the properties of multipartite entanglement in field theory. Even in analogous and much simpler qubit systems, multipartite entanglement isn't fully understood for systems of n qubits. The main emphasis of the AMPS argument is that the entanglement required between early and late radiation prevents entanglement across the horizon due to quantum monogamy and no cloning theorems. Quantum monogamy is simple to state for maximal entanglement for a bipartite pure state (e.g. a Bell pair). However these considerations are not completely general. For non-maximal multipartite entanglement for mixed states, quantum monogamy inequalities \cite{Wootters1998a} exist for various entanglement measures which limit entanglement, but do not strictly forbid the entanglement we are interested in.

One approach to the firewall question is to ask what is expected of the final state of the hawking radiation. The relevance of entanglement observed in this final state to the firewall problem can be seen by time evolving back the radiation's evolution to when it was just leaving the horizon. The question can be posed as what properties of the final state's entanglement say about the state of the quantum fields near the horizon region at an earlier time. Inspired by the discussion in \cite{Vidal2002}, we ideally might want to translate the question of how to restore the vacuum near a black hole's horizon into language that addresses, for example, the extent to which pure state entanglement can be extracted from infinitely many copies of the state. A concrete question like this would allow us to choose measures that have a given physical interpretation such as ``entanglement of formation'', ``distillable entanglement'', etc. which could illuminate further the nature of the vacuum.  However, the above translation is not straightforward and these measures are generally difficult to calculate.

\section{Conclusions} \label{sec:Conclusions}

In our toy model we constructed a history of states to model the evolution of a black hole. The method of state construction has ensured the state of the whole system remains pure, and enforced the evaporation process to only involve {\em local} unitary interactions. By introducing a multipartite entanglement measure, entanglement sharing among black hole and different parts of radiation can be measured. Our analysis has shown entanglement between the black hole and the qubit just radiated from the black hole for the entire history of evaporation. This differs from the AMPS's expectation that after the Page time, entanglement among these systems should be forbidden. 

Our very simple toy model explicitly shown entanglement across the horizon region isn't strictly forbidden. However, it is unclear how this result translates into an actual field theory description. Because of the limitations of our toy model, we cannot claim at this point that we know that a more realistic model could have enough entanglement or even the right type of entanglement to restore the expected properties of the field theory vacuum near the black hole. We also do not explicitly offer a scheme to restore the vacuum's purity which may be the more important property of the vacuum than the entanglement. This paper is meant to show a possible loophole in the AMPS argument. For a complete and proper treatment, we would want to extend this toy model to one more like a field theory. With a more realistic model, we could imagine time evolving back the hawking radiation and identifying what types of states you would have near the horizon for an infalling observer, thus potentially achieving insights into the firewall problem.

\section{Acknowledgments}
We thank Veronica Hubeny, Massimiliano Rota, Don Page, Rajiv Singh, Steven Carlip and William K. Wooters for helpful discussions. This work was supported in part by DOE grant DE-SC$0009999$.

\appendix

\section{}
In these Appendices we address several topics supporting the main points of the paper.  Appendices \ref{sec:negativity} and \ref{sec: firewall model table} define and develop the ideas of negativity, and Appendix \ref{sec:time evo} gives detailed information about our toy model.
\section{Negativity}\label{sec:negativity}

Negativity is an entanglement monotone meaning that it does not increase under local operations and classical communication. As discussed in \cite{Dur2000}, negativity may thus be considered an entanglement measure. As an illustration we can construct mixed state systems by first considering a pure state in a Hilbert space which is a product of three smaller spaces:  $U = H_{A}\otimes H_{B}\otimes H_{C}$. Tracing out system C, leaves density matrix $\rho_{A \otimes B}$.The negativity of $A,B$ for the state, $\rho_{A \otimes B}$ is given by:
\begin{equation}\label{eq:defneg}
\varepsilon_{A,B}(\rho_{A \otimes B})\equiv \frac{\parallel \rho^{T_{A}}_{A \otimes B} \parallel-1}{2} 
\end{equation}
where the trace norm $\parallel \rho^{T_{A}}_{A \otimes B} \parallel$ is the sum of the absolute values of the eigenvalues $\lambda_{i}$ of $\rho^{T_{A}}_{B}$. $\rho^{T_{A}}_{B}$ is the partial transpose of $B$ respect to $A$. The Peres - Horodecki criterion states that to have a state separable it is necessary the partial transpose of $\rho$ has only non-negative eigenvalues\cite{1996PhRvL..77.1413P}. If the partial transpose of $\rho$ has any negative eigenvalues then the state is necessarily inseparable. The trace norm measures how much $\rho^{T_{A}}_{A \otimes B}$ fails to be positive\cite{1996PhRvL..77.1413P}. Therefore if negativity is nonzero, then there is definitely entanglement. However, negativity being zero does not imply no entanglement. There exist a class of entangled states with zero negativity said to be PPT bound entangled states\cite{Vidal2002}. Thus it can be difficult to extract precise physical meaning from the measured value of negativity. It has been noted that negativity places a bound on the degree to which a single copy of the state $\rho$ can be used to perform quantum teleportation together with local operations and classical communications\cite{Vidal2002, Nielsen1999}. For our purposes in this paper we are mostly interested in its ability to identify when entanglement is present.

\section{Table of negativity for a man-made firewall model}\label{sec: firewall model table}

\begin{table}[h!]\label{table:3}
\centering
\begin{tabular}{ |p{2cm}||p{2cm}|p{2cm}|p{2cm}|  }
 \hline
 \multicolumn{4}{|c|}{Negativities in the Firewall Model} \\
 \hline
 Time & BH and JR &JR and OR&BH and OR\\
 \hline

		3 & 0.5 & 0.0 & 0.5\\
		4 & 0.5 & 0.0 & 0.5\\
		5 & 0.0 & 0.5 & 0.0\\

 \hline
\end{tabular}
\caption{Negativity values between all combinations of our three subsystems $BH$, $JR$ and $OR$ for all times. There is no entanglement between $BH$ and $JR$ at 5 because our choices of states are known to behave as so. Thus there is firewalls for all time when negativity is defined including after page time. The dashes implies the non-existed negativity values. For example, in the first time, since there is no subsystem $JR$ and $OR$, there are no negativity values between all combination of subsystems.} 
\end{table}

We have also calculated negativities for the Firewall toy model which is analogous to the first toy model in \cite{Stoltenberg2015}. This model is constructed with a final radiation state has each pair of qubits appearing in a Bell-pair. From the form of the states, it can be explicitly seen that entanglement exists between $BH$ and $JR$ until the Page Time (time 4), after which there is no entanglement present. This is consistent with the negativities shown in ~Table \ref{table:3}, where the negativity of $BH$ and $JR$ is non-zero prior to time 4 and then is zero after time 4. We remind the reader that trying gain more insight from comparing these values with ~Table\ref{table:2} can be misleading for the reasons listed in ~Sec.\ref{sec:Multipart}.

\section{Time evolution of the multipartite model} \label{sec:time evo}
Our method for generating time evolution is based on the change of basis
that we introduce once a qubit shifts from describing a black hole system to the radiation system. Section \ref{sec:toymodelsetup}, illustrates the way the new basis is used to construct the time evolution.
The choice of basis dictates what the entanglement will look like for the subsequent time state.
We wanted to create states that produced a $JR$ qubit that was maximally mixed as well as having non-zero
negativity between $JR$ and $BH$ (As explained in \ref{sec:toymodelsetup}). Finding a basis that accomplished this grew increasingly difficult as the states 
got more complicated and required some trial and error.
All mappings up till time 4 were done by hand, taking advantage of symmetries in the form of the states to find
the types of states that matched our criteria Getting states for time 5 and 6 were too difficult to do by hand, so the states were found by varying parameters in the basis vectors with a computer program.  The program is able to generate random orthonormal bases and do systematic mappings from one time to the next. Among the many mappings it generated, the program can select the set of states that have interesting negativities and have all the conditions we want to enforce to assure the evolution to be physical.\footnote{Since we generated the random bases for times 5 and 6 numerically, the eigenvalues of the density matrix of $JR$ for those times were not exactly $0.5$ but deviated from $0.5$ by less than $0.1$ percent. We do not expect this reflects any issue that would prevent cases with a Von Neumann entropy identical to unity in principle.}

In this section we give a detailed description of the toy model state at each of the discrete time steps $1$--$7$.  At each time the state is a pure state, thus ensuring unitary time evolution for the whole system.  Several key features have already been presented in Section~\ref{sec:toymodelsetup}.  These include the way we enforce locality by limiting which subsystems interact, and how we use basis mapping as a tool to generate the state at the $N^{th}$ time step from the state at the $(N-1)^{th}$ step.
The detailed process of the time evolution of the multipartite model is as follows:

The first time state is trivial, and is written
\begin{eqnarray}\label{eq:m1}
\ket{\psi_{BH, 1}}
\end{eqnarray}

The second time state is given by
\begin{eqnarray}\label{eq:m2}
\ket{\psi_{BH, 2}}= \frac{1}{\sqrt{2}}(\ket{0_{BH, 2}}\ket{0_{JR, 2}} \\
+ \ket{1_{BH, 2}}\ket{1_{JR, 2}})\nonumber
\end{eqnarray}
where  $\ket{0_{BH,2}}$ is the $0^{th}$ basis state for the black hole subsystem at time 2,etc.
Notice that there is nothing in OR subsystem at time 2.

The $i^{th}$ black hole basis state $\ket{i_{BH,2}}$ at time step 2 evolve into states that describe the black hole ($BH,3$) and the newly emitted particle ($JR,3$) at time step 3. This is done by writing the basis for ($BH,2$) in terms of ($BH,3$) 
\begin{eqnarray}\label{eq:m3}
\ket{0_{BH, 2}}=a\ket{0_{{BH, 3}}0_{{JR, 3}}}+a\ket{2_{{BH, 3}}1_{{JR, 3}}} \nonumber\\
+ b\ket{4_{{BH, 3}}0_{{JR, 3}}}\\
\ket{1_{BH, 2}}=a\ket{1_{{BH, 3}}0_{{JR, 3}}}+a\ket{3_{{BH, 3}}1_{{JR, 3}}} \nonumber\\
+ b\ket{4_{{BH, 3}}1_{{JR, 3}}}\nonumber
\end{eqnarray}
where the values of coefficients are assigned to be

\begin{eqnarray}\label{eq:coe}
a=\frac{1}{\sqrt{2}}\nonumber\\
b=0.\nonumber\\
\end{eqnarray}
We plug in this unitary change of basis into the state we had previously written to get the third time state:

\begin{eqnarray}\label{eq:m4}
\ket{\psi_{2}}=\frac{1}{\sqrt{2}}[(a\ket{0_{BH, 3}0_{JR, 3}}+a\ket{2_{BH, 3}1_{JR, 3}} \nonumber\\
+ b\ket{4_{BH, 3}0_{JR, 3}})\ket{0_{JR, 2}}\\ 
+ (a\ket{1_{BH, 3}0_{JR, 3}}+a\ket{3_{BH, 3}1_{JR, 3}}\nonumber\\
+ b\ket{4_{BH, 3}1_{JR, 3}})\ket{1_{JR, 2}})].\nonumber
\end{eqnarray}
%
%
%

The mapping we use to go from time $3$ to time $4$ is 
\begin{eqnarray}\label{eq:m6}
\ket{0_{BH, 3}}=a_{1}(\ket{0_{{BH, 4}}1_{{JR, 4}}}+\ket{1_{{BH, 4}}0_{{JR, 4}}}) \nonumber\\+ b_{1}(\ket{2_{{BH, 4}}0_{{JR, 4}}}-\ket{3_{{BH, 4}}1_{{JR, 4}}})\nonumber\\
\ket{1_{BH, 3}}=a_{1}(\ket{4_{{BH, 4}}0_{{JR, 4}}}+\ket{5_{{BH, 4}}1_{{JR, 4}}}) \nonumber\\+ b_{1}(-\ket{6_{{BH, 4}}0_{{JR, 4}}}+\ket{7_{{BH, 4}}1_{{JR, 4}}})\nonumber\\
\ket{2_{BH, 3}}=a_{1}(\ket{4_{{BH, 4}}0_{{JR, 4}}}-\ket{5_{{BH, 4}}1_{{JR, 4}}}) \\+ b_{1}(\ket{6_{{BH, 4}}0_{{JR, 4}}}+\ket{7_{{BH, 4}}1_{{JR, 5}}})\nonumber\\
\ket{3_{BH, 3}}=a_{1}(\ket{0_{{BH, 4}}1_{{JR, 4}}}-\ket{1_{{BH, 4}}0_{{JR, 4}}}) \nonumber\\+ b_{1}(\ket{2_{{BH, 4}}0_{{JR, 4}}}+\ket{3_{{BH, 4}}1_{{JR, 5}}})\nonumber\\
\ket{4_{BH, 3}}=a_{1}(\ket{2_{{BH, 4}}0_{{JR, 4}}}\ket{3_{{BH, 4}}1_{{JR, 4}}}) \nonumber\\+ b_{1}(-\ket{0_{{BH, 4}}1_{{JR, 4}}}+\ket{1_{{BH, 4}}0_{{JR, 4}}})\nonumber
\end{eqnarray}
where 
\begin{eqnarray}\label{eq:coe1}
a_{1} = 0.016\nonumber\\
b_{1} = 0.707\\
\end{eqnarray}

We plug this unitary change of basis into the time $3$ state to get the toy model state for time $4$.

Due to the increasing complexity of the expressions, for time steps $4$ and higher we only give the basis mappings, and do not explicitly present the outcome of the substitutions.  The information we do provide is sufficient to fully reproduce our results.

The mapping from time $4$ to time $5$ is 
\begin{widetext}
\begin{eqnarray}\label{eq:m8}
\ket{0_{BH, 4}}= c_{0,00}\ket{0_{{BH, 5}}0_{{JR, 5}}} + c_{0,01}\ket{0_{{BH, 5}}1_{{JR, 5}}}
+ c_{0,10}\ket{1_{{BH, 5}}0_{{JR, 5}}} + c_{0,11}\ket{1_{{BH, 5}}1_{{JR, 5}}}\nonumber\\
+ c_{0,20}\ket{2_{{BH, 5}}0_{{JR, 5}}} + c_{0,21}\ket{2_{{BH, 5}}1_{{JR, 5}}}
+ c_{0,30}\ket{3_{{BH, 5}}0_{{JR, 5}}} + c_{0,31}\ket{3_{{BH, 5}}1_{{JR, 5}}}\nonumber\\
\ket{1_{BH, 4}}= c_{1,00}\ket{0_{{BH, 5}}0_{{JR, 5}}} + c_{1,01}\ket {0_{{BH, 5}}1_{{JR, 5}}}
+ c_{1,10}\ket{1_{{BH, 5}}0_{{JR, 5}}} + c_{1,11}\ket{1_{{BH, 5}}1_{{JR, 5}}}\nonumber\\
+ c_{1,20}\ket{2_{{BH, 5}}0_{{JR, 5}}} + c_{1,21}\ket{2_{{BH, 5}}1_{{JR, 5}}}
+ c_{1,30}\ket{3_{{BH, 5}}0_{{JR, 5}}} + c_{1,31}\ket{3_{{BH, 5}}1_{{JR, 5}}}\nonumber\\
\ket{2_{BH, 4}}=c_{2,00}\ket{0_{{BH, 5}}0_{{JR, 5}}} + c_{2,01}\ket{0_{{BH, 5}}1_{{JR, 5}}}
+ c_{2,10}\ket{1_{{BH, 5}}0_{{JR, 5}}} + c_{2,11}\ket{1_{{BH, 5}}1_{{JR, 5}}}\nonumber\\
+ c_{2,20}\ket{2_{{BH, 5}}0_{{JR, 5}}} + c_{2,21}\ket{2_{{BH, 5}}1_{{JR, 5}}}
+ c_{2,30}\ket{3_{{BH, 5}}0_{{JR, 5}}} + c_{2,31}\ket{3_{{BH, 5}}1_{{JR, 5}}}\nonumber\\
\ket{3_{BH, 4}}=c_{3,00}\ket{0_{{BH, 5}}0_{{JR, 5}}} + c_{3,01}\ket{0_{{BH, 5}}1_{{JR, 5}}}
+ c_{3,10}\ket{1_{{BH, 5}}0_{{JR, 5}}} + c_{3,11}\ket{1_{{BH, 5}}1_{{JR, 5}}}\nonumber\\
+ c_{3,20}\ket{2_{{BH, 5}}0_{{JR, 5}}} + c_{3,21}\ket{2_{{BH, 5}}1_{{JR, 5}}}
+ c_{3,30}\ket{3_{{BH, 5}}0_{{JR, 5}}} + c_{3,31}\ket{3_{{BH, 5}}1_{{JR, 5}}}\nonumber\\
\ket{4_{BH, 4}}=c_{4,00}\ket{0_{{BH, 5}}0_{{JR, 5}}} + c_{4,01}\ket{0_{{BH, 5}}1_{{JR, 5}}}
+ c_{4,10}\ket{1_{{BH, 5}}0_{{JR, 5}}} + c_{4,11}\ket{1_{{BH, 5}}1_{{JR, 5}}}\nonumber\\
+ c_{4,20}\ket{2_{{BH, 5}}0_{{JR, 5}}} + c_{4,21}\ket{2_{{BH, 5}}1_{{JR, 5}}}
+ c_{4,30}\ket{3_{{BH, 5}}0_{{JR, 5}}} + c_{4,31}\ket{3_{{BH, 5}}1_{{JR, 5}}}\\
\ket{5_{BH, 4}}=c_{5,00}\ket{0_{{BH, 5}}0_{{JR, 5}}} + c_{5,01}\ket{0_{{BH, 5}}1_{{JR, 5}}}
+ c_{5,10}\ket{1_{{BH, 5}}0_{{JR, 5}}} + c_{5,11}\ket{1_{{BH, 5}}1_{{JR, 5}}}\nonumber\\
+ c_{5,20}\ket{2_{{BH, 5}}0_{{JR, 5}}} + c_{5,21}\ket{2_{{BH, 5}}1_{{JR, 5}}}
+ c_{5,30}\ket{3_{{BH, 5}}0_{{JR, 5}}} + c_{5,31}\ket{3_{{BH, 5}}1_{{JR, 5}}}\nonumber\\
\ket{6_{BH, 4}}=c_{6,00}\ket{0_{{BH, 5}}0_{{JR, 5}}} + c_{6,01}\ket{0_{{BH, 5}}1_{{JR, 5}}}
+ c_{6,10}\ket{1_{{BH, 5}}0_{{JR, 5}}} + c_{6,11}\ket{1_{{BH, 5}}1_{{JR, 5}}}\nonumber\\
+ c_{6,20}\ket{2_{{BH, 5}}0_{{JR, 5}}} + c_{6,21}\ket{2_{{BH, 5}}1_{{JR, 5}}}
+ c_{6,30}\ket{3_{{BH, 5}}0_{{JR, 5}}} + c_{6,31}\ket{3_{{BH, 5}}1_{{JR, 5}}}\nonumber\\
\ket{7_{BH, 4}}=c_{7,00}\ket{0_{{BH, 5}}0_{{JR, 5}}} + c_{7,01}\ket{0_{{BH, 5}}1_{{JR, 5}}}
+ c_{7,10}\ket{1_{{BH, 5}}0_{{JR, 5}}} + c_{7,11}\ket{1_{{BH, 5}}1_{{JR, 5}}}\nonumber\\
+ c_{7,20}\ket{2_{{BH, 5}}0_{{JR, 5}}} + c_{7,21}\ket{2_{{BH, 5}}1_{{JR, 5}}}
+ c_{7,30}\ket{3_{{BH, 5}}0_{{JR, 5}}} + c_{7,31}\ket{3_{{BH, 5}}1_{{JR, 5}}}\nonumber
\end{eqnarray}\\
\end{widetext}

Each of the term associate with a coefficient $c_{i,j}$. If coefficients c represented as matrix, the values are assigned as the following: 

\begin{widetext}
		\begin{align}
&	\begin{Bmatrix}
c_{0,00} & c_{1,00} & c_{2,00} & c_{3,00} & c_{4,00} & c_{5,00} &c_{6,00} & c_{7,00}\\
c_{0,01} & c_{1,01} & c_{2,01} & c_{3,01} & c_{4,01} & c_{5,01} &c_{6,01} & c_{7,01}\\
c_{0,10} & c_{1,10} & c_{2,10} & c_{3,10} & c_{4,10} & c_{5,10} &c_{6,10} & c_{7,10}\\
c_{0,11} & c_{1,11} & c_{2,11} & c_{3,11} & c_{4,11} & c_{5,11} &c_{6,11} & c_{7,11}\\
c_{0,20} & c_{1,20} & c_{2,20} & c_{3,20} & c_{4,20} & c_{5,20} &c_{6,20} & c_{7,20}\\
c_{0,21} & c_{1,21} & c_{2,21} & c_{3,21} & c_{4,21} & c_{5,21} &c_{6,21} & c_{7,21}\\
c_{0,30} & c_{1,30} & c_{2,30} & c_{3,30} & c_{4,30} & c_{5,30} &c_{6,30} & c_{7,30}\\
c_{0,31} & c_{1,31} & c_{2,31} & c_{3,31} & c_{4,31} & c_{5,31} &c_{6,31} & c_{7,31}\\
	\end{Bmatrix} \nonumber
\\
&\ \ \ \ \ \ \ \ \ \ \ \ \ \ \ \ \ \ \ \ \ \ \ \ \ \ \ \ \ \ \ \ \ \ \ \|\nonumber
\\
&	\begin{bmatrix}
0.3552504 & -0.3231891 & -0.0253468 & -0.2101122 & 0.0047422 & 0.251555 & -0.0296075 & -0.8126361\\
-0.019072 & 0.1915158 & -0.4876553 & -0.2481800 & -0.0335412 &  0.5521110 & -0.5682440 & 0.1862906\\
-0.0586458 & 0.7537837 & -0.1624502 & -0.106498 & 0.3785648 & -0.312295 & -0.0106699 & -0.3868923\\
0.5933005 & -0.1305636 & -0.5811510 & 0.117670 & -0.1202273 & -0.4915727 & -0.0387296 & 0.147535\\
0.1742461 & 0.065539 & -0.0619712 & -0.7366273 & 0.0339155 & 0.1125694 & 0.5825466 & 0.2563204\\
0.2562844 & 0.1236680 & 0.5342872 & -0.4165860 & -0.2992278 & -0.3197298 & -0.5129216 & 0.0718675\\
-0.5052435 & 0.0170885 & -0.2934006 & -0.167775 & -0.7153596 & -0.2153753 & 0.0989234 & -0.2495859\\
-0.4082575 & -0.5037386 & -0.1484463 & -0.353213 & 0.4885280 & -0.3625415 & -0.2497468 & 0.0175458\\
	\end{bmatrix}
		\end{align}
\end{widetext}
    
The expression of mapping from time $5$ to time $6$ is similar. The coefficients matrix for mapping from time $5$ to time $6$ is:

\begin{widetext}
\begin{eqnarray}
& \begin{bmatrix}
~^1\!c_{0,00} & ~^1\!c_{1,00} & ~^1\!c_{2,00} & ~^1\!c_{3,00}\\
~^1\!c_{0,01} & ~^1\!c_{1,01} & ~^1\!c_{2,01} & ~^1\!c_{3,01}\\
~^1\!c_{0,10} & ~^1\!c_{1,10} & ~^1\!c_{2,10} & ~^1\!c_{3,10}\\
~^1\!c_{0,11} & ~^1\!c_{1,11} & ~^1\!c_{2,11} & ~^1\!c_{3,11}\\
\end{bmatrix} \nonumber
\\
&\ \ \ \|\nonumber 
\\
& \begin{bmatrix}
0.5840557 & -0.7331548 & 0.172482 & 0.3026765\\
0.4858683 & 0.1387961 & 0.3380292 & -0.7939798\\
0.262656 & 0.5607161 & 0.5952301 & 0.5121625\\
-0.5948299 & -0.3589111 & 0.7082990 & -0.1251904
\end{bmatrix}
\end{eqnarray}
\end{widetext}

The final step from $6$ to $7$ is trivial, in that the single $BH$ qubit is reassigned to $JR$, and the old $JR$ qubit is reassigned to $OR$.  No black hole remains.

\bibliography{july22ref}

\end{document}